\documentclass{article}
\usepackage{amssymb}
\usepackage{amsmath}

\input{tcilatex}
\begin{document}

\title{Comment on " Approximate Analytical Versus Numerical Solutions of Schr%
\"{o}dinger Equation Under Molecular Hua Potential "}
\author{M. Ferdjaoui, A. Khodja, F. Benamira and L. Guechi* \\
Laboratoire de Physique Th\'{e}orique, D\'{e}partement de Physique, \and %
Facult\'{e} des Sciences Exactes, Universit\'{e} des fr\`{e}res Mentouri,
\and Constantine 1, Route d'Ain El Bey, Constantine, Algeria \and *
Corresponding author: guechilarbi@yahoo.fr}
\maketitle

\begin{abstract}
We present arguments proving that the results obtained by Hassanabadi and
coworkers \cite{Hassanabadi} in the study of the D-dimensional Schr\"{o}%
dinger equation with molecular Hua potential through the supersymmetry
method in quantum mechanics are incorrect. We identified the inconsistencies
in their reasoning on the allowed values of the parameter $q$ and we
constructed the correct energy spectrum.
\end{abstract}

About a decade ago Hassanabadi and his coworkers \cite{Hassanabadi} claimed
to have approximately solved the D-dimensional Schr\"{o}dinger equation with
the Hua potential in the framework of the supersymmetric quantum mechanics
approach (SUSY QM) by employing a Pekeris-type approximation to replace the
centrifugal potential term. We point out however that there are several
inconsistencies in the application of the SUSY QM and in the derivation of
the energy spectrum.

First, the Hua potential \cite{Hua} is given by the expression%
\begin{equation}
V(r)=V_{0}\left[ \frac{1-e^{-b_{h}(r-r_{e})}}{1-qe^{-b_{h}(r-r_{e})}}\right]
^{2}  \label{E.1}
\end{equation}%
with the deformation parameter $q$ contained in the interval $-1<q<1$. For $%
q>0$, it is obvious that the potential (\ref{E.1}) has a strong singularity
at the point $r=r_{0}=r_{e}+\frac{1}{b_{h}}\ln q$, and on the other hand the
Pekeris approximation%
\begin{equation}
\frac{1}{r^{2}}\approx \frac{1}{r_{e}^{2}}\left[ D_{0}+D_{1}\frac{%
e^{-b_{h}(r-r_{e})}}{1-qe^{-b_{h}(r-r_{e})}}+D_{2}\frac{e^{-2b_{h}(r-r_{e})}%
}{\left( 1-qe^{-b_{h}(r-r_{e})}\right) ^{2}}\right]  \label{E.2}
\end{equation}%
is valid only for $qe^{b_{h}r_{e}}\geq 1$ (see Refs. \cite{Benamira,Khodja1}%
). The D-dimensional Schr\"{o}dinger equation (4) in Ref. \cite{Hassanabadi}
should be written in the range $r_{0}<r<\infty $ and for $%
e^{-b_{h}r_{e}}\leq q<1$, as

\begin{eqnarray}
&&\left\{ -\frac{d^{2}}{dr^{2}}+\frac{2\mu }{\hbar ^{2}}V(r)+A_{l}\left[
D_{0}+\frac{D_{1}e^{-b_{h}(r-r_{e})}}{1-qe^{-b_{h}(r-r_{e})}}+\frac{%
D_{2}e^{-2b_{h}(r-r_{e})}}{\left( 1-qe^{-b_{h}(r-r_{e})}\right) ^{2}}\right]
\right\} R_{n_{r},l}(r)  \notag \\
&=&\frac{2\mu }{\hbar ^{2}}ER_{n_{r},l}(r),  \label{E.3}
\end{eqnarray}%
where $A_{l}=\frac{\left( D+2l-1\right) \left( D+2l-3\right) }{4r_{e}^{2}}.$

Second, with the superpotential $\phi (r)$ defined as

\begin{equation}
\phi (r)=\frac{A}{1-qe^{-\alpha x}}+B;\text{ }x=\frac{r-r_{e}}{r_{e}}\text{
and }\alpha =b_{h}r_{e},  \label{E.4}
\end{equation}%
the authors of Ref. \cite{Hassanabadi}\ obtained the Riccati equation

\begin{equation}
\phi ^{2}(r)-\phi ^{\prime }(r)=V_{eff}(r)-\widetilde{E}_{0,l},  \label{E.5}
\end{equation}%
from which the quantities $A,B$ and $\widetilde{E}_{0,l}$ are found to be%
\begin{equation}
A=-\frac{\alpha }{2}\left[ 1\pm \sqrt{1-\frac{4V_{1}}{\alpha ^{2}}}\right] ,
\label{E.6a}
\end{equation}%
\begin{equation}
B=\frac{\alpha }{2}\left[ 1-\frac{V_{2}}{\alpha A}\right] ,  \label{E.6b}
\end{equation}%
\begin{equation}
\widetilde{E}_{0,l}=-B^{2}.  \label{E.6c}
\end{equation}%
Without correctly specifying the signs of $A$ and $B$, they then used the
shape invariance approach to obtain the energy spectrum. Therefore, the
result given by Eq. (21) in Ref. \cite{Hassanabadi} is not correct. In this
case, the signs of $A$ and $B$ can be fixed by considering the ground state
wave function $R_{0,l}(r)$ defined by

\begin{equation}
R_{0,l}(r)=Ne^{-\int \phi \left( x\right) dx}=\mathcal{N}e^{-\left(
A+B\right) x}\left( 1-qe^{-\alpha x}\right) ^{-\frac{A}{\alpha }},
\label{E.7}
\end{equation}%
where $\mathcal{N}$ is the normalisation constant. For $R_{0,l}(r)$ to be a
physically acceptable solution, it has to satisfy the boundary conditions

\begin{equation}
R_{0,l}(r)\underset{r\rightarrow \infty }{\rightarrow }0,  \label{E.8}
\end{equation}%
and

\begin{equation}
R_{0,l}(r)\underset{r\rightarrow r_{0}}{\rightarrow }0.  \label{E.9}
\end{equation}%
From this we see that $A<0$ and $A+B>0$ or $B>\left\vert A\right\vert $. The
solution of the problem should be re-examined starting from the resolution
of equations (15a) and (15b) in Ref. \cite{Hassanabadi}\ . As a result, $A$
and $B$ can be expressed as%
\begin{equation}
A=-\frac{\alpha }{2}\left[ 1+\sqrt{1-\frac{4V_{1}}{\alpha ^{2}}}\right] ,
\label{E.10a}
\end{equation}%
\begin{equation}
B=-\frac{1}{2}\left( A+\frac{V_{1}+V_{2}}{A}\right) .  \label{E.10b}
\end{equation}%
Then, by putting $a_{0}=A$ and using the shape invariance condition%
\begin{equation}
V_{eff\text{ }+}\left( x,a_{0}\right) =V_{eff\text{ }-}\left( x,a_{1}\right)
+R\left( a_{1}\right) ,  \label{E.11}
\end{equation}%
we find after some simple calculation that%
\begin{equation}
R\left( a_{1}\right) =\frac{1}{4}\left[ \left( a_{0}+\frac{V_{1}+V_{2}}{a_{0}%
}\right) ^{2}-\left( a_{1}+\frac{V_{1}+V_{2}}{a_{1}}\right) ^{2}\right] ,
\label{E.12}
\end{equation}%
and%
\begin{equation}
a_{1}=a_{0}-\alpha .  \label{E.13}
\end{equation}%
The energy eigenvalues of Hamiltonian $H_{-}=-\frac{d^{2}}{dx^{2}}+V_{eff%
\text{ }-}\left( x\right) $ are then given by%
\begin{equation}
\widetilde{E}_{n_{r},l}^{\left( -\right) }=\underset{k=1}{\overset{n_{r}}{%
\sum }}R\left( a_{k}\right) =\frac{1}{4}\left[ \left( a_{0}+\frac{V_{1}+V_{2}%
}{a_{0}}\right) ^{2}-\left( a_{0}-n_{r}\alpha +\frac{V_{1}+V_{2}}{%
a_{0}-n_{r}\alpha }\right) ^{2}\right] .  \label{E.14}
\end{equation}%
From Eqs. (\ref{E.6c}) and (\ref{E.14}) it follows immediately that%
\begin{equation}
\widetilde{E}_{n_{r},l}=\widetilde{E}_{n_{r},l}^{\left( -\right) }+%
\widetilde{E}_{0,l}=-\frac{1}{4}\left( a_{0}-n_{r}\alpha +\frac{V_{1}+V_{2}}{%
a_{0}-n_{r}\alpha }\right) ^{2},  \label{E.15}
\end{equation}%
By using Eq. (\ref{E.10a}) together with Eqs. (10) in the Ref. \cite%
{Hassanabadi} and since $\widetilde{E}_{n_{r},l}=V_{3}$ (see Eq. (12) in
Ref. \cite{Hassanabadi}) we arrive at the following expression for the
energy levels:%
\begin{eqnarray}
E_{n_{r},l} &=&\frac{V_{0}}{2}\left( 1+\frac{1}{q^{2}}\right) -\frac{\hbar
^{2}b_{h}^{2}}{8\mu }\left[ N_{r}^{2}+\frac{\lambda _{l}^{2}}{N_{r}^{2}}%
\right]  \notag \\
&&+\frac{\hbar ^{2}}{8\mu r_{e}^{2}}\left( D+2l-1\right) \left(
D+2l-3\right) \left[ D_{0}+\frac{1}{2q}\left( \frac{D_{2}}{q}-D_{1}\right) %
\right] ;\text{ \ }e^{-b_{h}r_{e}}\leq q<1,  \notag \\
&&  \label{E.16}
\end{eqnarray}%
where we have set%
\begin{equation}
N_{r}=n_{r}+\delta _{l}+\frac{1}{2},\text{ \ \ \ \ \ }\lambda _{l}=\frac{%
2\mu V_{0}}{\hbar ^{2}b_{h}^{2}}\left( \frac{1}{q^{2}}-1\right) +\frac{%
\left( D+2l-1\right) \left( D+2l-3\right) }{4b_{h}^{2}r_{e}^{2}}\left( \frac{%
D_{2}}{q^{2}}-\frac{D_{1}}{q}\right) ,\text{\ \ }  \label{E.17}
\end{equation}%
and%
\begin{equation}
\delta _{l}=\sqrt{\frac{1}{4}+\frac{2\mu V_{0}}{\hbar ^{2}b_{h}^{2}}\left( 1-%
\frac{1}{q}\right) ^{2}+\frac{\left( D+2l-1\right) \left( D+2l-3\right) }{%
4b_{h}^{2}r_{e}^{2}}\frac{D_{2}}{q^{2}}\text{\ }}.  \label{E.18}
\end{equation}%
This result can be verified in three-dimensional space. Indeed, if one
substitutes $D_{0}=C_{0}r_{e}^{2},D_{1}=B_{0}r_{e}^{2},D_{2}=A_{0}r_{e}^{2}$%
, and $q=c_{h},$ one recovers the discrete energy spectrum derived by path
integration \cite{Khodja2}.

Third, the numerical results obtained from Eq. (21) in the Ref. \cite%
{Hassanabadi} for $r_{e}=1,q=0.170066$ and $b_{h}=1.61890$ in table $1$, are
wrong. In this case, the correct numerical values must be calculated from
the expression of our Eq. (\ref{E.16}) which is valid for $q\geq 0.198116507$
when $r_{e}=1$ and $b_{h}=1.61890$. We can also point out that the variation
of $E_{n_{r},0}$ in terms of the parameter $q$ is valid only for $%
e^{-b_{h}r_{e}}\leq q<1$ (see Fig. (6) in Ref. \cite{Hassanabadi} ). In
addition, when $q=1$, the potential (\ref{E.1}) becomes a step potential for
which there are no bound states. This makes it possible to affirm that the
curves plotted by the authors of Ref. \cite{Hassanabadi} in Fig. (6) are
incorrect.

In conclusion, the approximate analytical and numerical results obtained by
the authors of Ref. \cite{Hassanabadi} are unsatisfactory because \ the SUSY
QM method is used without taking into account the conditions for its
application. The radial Schr\"{o}dinger equation (\ref{E.3}) can only be
approximately solved by this method when $e^{-b_{h}r_{e}}\leqslant c_{h}<1$
and $r_{0}<r<+\infty $.

\end{document}